\documentclass[useAMS,usegraphicx,usenatbib]{mn2e}
\usepackage{graphicx}
\usepackage{amsmath}
\usepackage{amssymb}
\usepackage{subfigure}
\usepackage{url}
\usepackage{multirow}
\usepackage[linkcolor=blue,colorlinks=true,breaklinks,citecolor=blue,urlcolor=blue]{hyperref} 

\newcommand{\cm}{\rm\thinspace cm}

\newcommand{\s}{\rm\thinspace s}
\newcommand{\ks}{\rm\thinspace ks}

\newcommand{\Ms}{\rm\thinspace Ms}

\newcommand{\Hz}{\rm\thinspace Hz}

\newcommand{\keV}{\rm\thinspace keV}

\newcommand{\erg}{\rm\thinspace erg}

\newcommand{\ergpssq}{\hbox{$\erg\s^{-2}\,$}}

\newcommand{\cts}{\rm\thinspace ct}
\newcommand{\ctsps}{\hbox{$\cts\s^{-1}\,$}}

\newcommand{\ergcmps}{\hbox{$\erg\cm\ps\,$}}

\newcommand{\asec}{\rm\thinspace arcsec}

\newcommand{\ps}{\hbox{$\s^{-1}\,$}}

\newcommand{\rg}{\rm\thinspace $r_\mathrm{g}$}

\title[The changing corona of 1H\,0707$-$495]{Caught in the act: Measuring the changes in the corona that cause the extreme variability of 1H\,0707$-$495}
\author[D. R. Wilkins \textit{et al.}]{D. R. Wilkins$^{1,2}$
  \thanks{E-mail: drw@ap.smu.ca}\thanks{CITA National Fellow}, E. Kara$^{2}$, A. C. Fabian$^{2}$ and L. C. Gallo$^{1}$\\$^1$Department of Astronomy \& Physics, Saint Mary's University, Halifax, NS. B3H 3C3 Canada\\$^2$Institute of Astronomy, University of Cambridge, Madingley Road, Cambridge. CB3 0HA UK}
\begin{document}

\date{Accepted 2014 June 24.  Received 2014 June 22; in original form 2014 February 19}

\pagerange{\pageref{firstpage}--\pageref{lastpage}} \pubyear{2014}

\maketitle

\label{firstpage}

\begin{abstract}
The X-ray spectra of the narrow line Seyfert 1 galaxy, 1H\,0707$-$495, obtained with \textit{XMM-Newton}, from time periods of varying X-ray luminosity are analysed in the context of understanding the changes to the X-ray emitting corona that lead to the extreme variability seen in the X-ray emission from active galactic nuclei (AGN). The emissivity profile of the accretion disc, illuminated by the X-ray emitting corona, along with previous measurements of reverberation time lags are used to infer the spatial extent of the X-ray source. By fitting a twice-broken power law emissivity profile to the relativistically-broadened iron K$\alpha$ fluorescence line, it is inferred that the X-ray emitting corona expands radially, over the plane of the accretion disc, by 25 to 30 per cent as the luminosity increases, contracting again as the luminosity decreases, while increases in the measured reverberation lag as the luminosity increases would require also variation in the vertical extent of the source above the disc. The spectrum of the X-ray continuum is found to soften as the total X-ray luminosity increases and we explore the variation in reflected flux as a function of directly-observed continuum flux. These three observations combined with simple, first-principles models constructed from ray tracing simulations of extended coron\ae\ self-consistently portray an expanding corona whose average energy density decreases, but with a greater number of scattering particles as the luminosity of this extreme object increases.
\end{abstract}


\section{Introduction}
The X-ray emission from the accreting black holes in active galactic nuclei (AGN) is highly variable, particularly in narrow line Seyfert 1 (NLS1) galaxies \citep{leighly-99_2,turner+99}. For instance, the X-ray count rate observed from the NLS1 galaxy 1H\,0707$-$495 is seen to vary by factors of two to three on timescales of just a few hours \citep{fabian+09} while \citet{iras_fix} find that the X-ray emission from the NLS1 galaxy IRAS\,13224$-$3809 rises by a factor of three on timescales of just half an hour; an increase at a rate of $7\times 10^{40}$\ergpssq and even decrease by 50 per cent in only a few hundred seconds; a rate approaching $10^{42}$\ergpssq.

Detailed studies of this X-ray emission and, particularly, its variability have recently added a further dimension to the study of accreting black holes. X-rays are emitted from a corona of energetic particles surrounding the central black hole. These particles are thought to be accelerated and confined by magnetic fields arising from the ionised accretion disc \citep{galeev+79,haardt+91,merloni_fabian} and they inverse-Compton scatter thermal seed photons from the accretion disc to the X-ray energies that are observed with a power law spectrum \citep{sunyaev_trumper}.

In addition to being observed directly, this coronal X-ray continuum is seen to be reflected from the optically thick, geometrically thin accretion disc \citep{george_fabian}. X-rays incident on the accretion disc are backscattered, and fluorescent lines as well as secondary emission caused by heating of the gas are produced \citep{fabian_ross_rev}. The most prominent of these lines is the K$\alpha$ fluorescence line of iron at 6.4\keV\ for neutral iron \citep{matt+97}. This emission line is seen clearly in the X-ray spectra of many accreting black holes and is broadened by relativistic effects between the emitting material in the accretion disc and the observer \citep{fabian+89}; the combination of Doppler shifts and relativistic beaming from the orbital motion of the material and the redshift from the strong gravitational field close to the black hole give the line a characteristic blue shifted `horn' and extended redshifted `wing' to low energies.

\citet{1h0707_emis_paper} find that the profile of this relativistically broadened emission line reveals the \textit{emissivity profile}, that is the radial illumination pattern of the accretion disc by the coronal X-ray source. This emissivity profile is, in turn, sensitive to the spatial extent of the X-ray emission allowing the location and geometry of the X-ray emitting corona to be constrained by observational data. In 1H\,0707$-$495, the corona was found to extend radially outwards over the accretion disc to around 35\rg\ \citep{understanding_emis_paper}, where a gravitational radius, 1\rg$={GM}/{c^2}$, is the characteristic scale-length in the gravitational field around a point mass such as the black hole while extending just a couple of gravitational radii above the plane of the accretion disc \citep{lag_spectra_paper}.

It is through the reflection of X-rays from the accretion disc that the variability of the emission can be exploited as a probe of the innermost regions. Due to the additional path the light must travel between the corona and accretion disc, variability in the X-rays reflected from the disc is seen to lag behind the corresponding variability in the directly observed continuum emission that is illuminating it. \citet{fabian+09} and \citet{zoghbi+09} first measured these so-called reverberation time lags in 1H\,0707$-$495 and found them to correspond to the light travel time over around two gravitational radii, indicating that the reflected component of the X-ray spectrum really is being emitted from the innermost regions of the accretion flow around the black hole and allowing the structure of the corona and accretion flow to be probed using these measurements. Analogous reverberation time lags have since been measured in a multitude of further AGN \citep[\textit{e.g.}][]{emmanoul+2011,demarco+2011,zoghbi+2011,zoghbi+2012,demarco+2012}. Measuring the reverberation time lag as a function of energy reveals the redshifted wing of the broad iron K$\alpha$ emission line emitted from the innermost part of the accretion disc (and hence closer to the coronal emission) responding to changes in the continuum before emission from the outer disc \citep{zoghbi+2012,kara+12}.

Variations in luminosity are widely attributed to fluctuations in the mass accretion rate onto the black hole, with over- and under-densities in the accretion flow that diffuse inwards on viscous timescales in the disc \citep[\textit{e.g.}][]{kotov+2001,arevalo+2006}. The gravitational binding energy liberated by the accretion process is therefore modulated by the fluctuations in the density of the accretion disc, causing the luminosity we observe to fluctuate.

In order to understand the exact cause of this variability and also the underlying process by which energy is liberated from the accretion flow to power the X-ray emission that is observed, the X-ray spectra of accreting black holes have been studied, in detail, at both high and low flux levels. \citet{fabian_vaughan} compare the X-ray spectrum of another NLS1 galaxy, MCG--6-30-15 in states of high and low flux, finding that the flux reflected from the accretion disc remains almost constant while it is the flux in the continuum emission that varies. This apparent constancy of the reflected flux while the illuminating continuum varies can be understood in terms of a compact coronal X-ray source moving closer to and further from the black hole \citep{miniutti+03}. As the X-ray source moves closer to the black hole, more photons are focussed towards the black hole and on to the inner parts of the accretion disc, which means fewer photons are able to escape to be observed in the continuum, while the reduction in continuum flux is compensated by more photons being focussed on to the disc, causing the reflected flux to appear constant.

We here discuss detailed analysis of the X-ray spectra of the NLS1 galaxy 1H\,0707$-$495 from periods of varying luminosity. Combining the profile of the relativistically broadened iron K$\alpha$ emission line, the spectrum of the continuum emission and the variation in reflected and directly observed fluxes, along with simple models, we trace the underlying changes to corona that are driving the extreme variability we observe in the X-ray emission.

\subsection{Measuring the Corona through Reverberation}
\citet{kara_iras_lags} obtain the lag spectrum of the 0.3-1.0\keV\ band, dominated by reflection from the accretion disc, relative to the 1.0-4.0\keV\ band, dominated by the directly observed continuum emitted from corona during periods in which the total observed flux from a different NLS1 galaxy, IRAS\,13224$-$3809 is low, compared to the periods in which it flares. The lag spectrum shows the time delay between correlated variability in the two energy bands. They find that the reverberation lag between variability in the `primary' band and that in the `reflected' band increases from around 230\s\ in the low flux state to around 660\s\ in the high flux state. The Fourier frequency at which the lag is seen decreases from $4.1\times 10^{-4}$\Hz\ to $1.8\times 10^{-4}$\Hz.

Interpreted in the context of X-rays emitted from a corona around the central black hole and reflecting off the accretion disc, such a result is due to the average light travel time between the source and the reflector increasing. Looking at theoretical lag spectra obtained in ray tracing simulations \citep[see, \textit{e.g.},][]{lag_spectra_paper}, it can be seen that while increasing the radial extent of the coronal X-ray source decreases the observed lag time, increasing the vertical extent of the X-ray source increases the measured lag. This is not to say that the radial extent of the X-ray source is not also increasing, merely that the dominant effect is due to the increasing vertical extent of the source. It should be noted that like-for-like, the extra lag time due to increasing the vertical extent of the source is, in itself, a greater effect than that of increasing the radial extent of the source since the vertical extent of the source acts to extend the light path to the disc, while increasing the radial extent of the source decreases the impact of the Shapiro delay close to the black hole as more rays are now travelling further from the black hole to reach the accretion disc.

Analysis of reverberation lags through lag spectra requires continuous light curves. It is therefore only possible to compare the lag spectra during low and high flux states when there are extended, continuous periods during which the source can be found in such a state.

\begin{figure*}
\centering
\includegraphics[width=170mm]{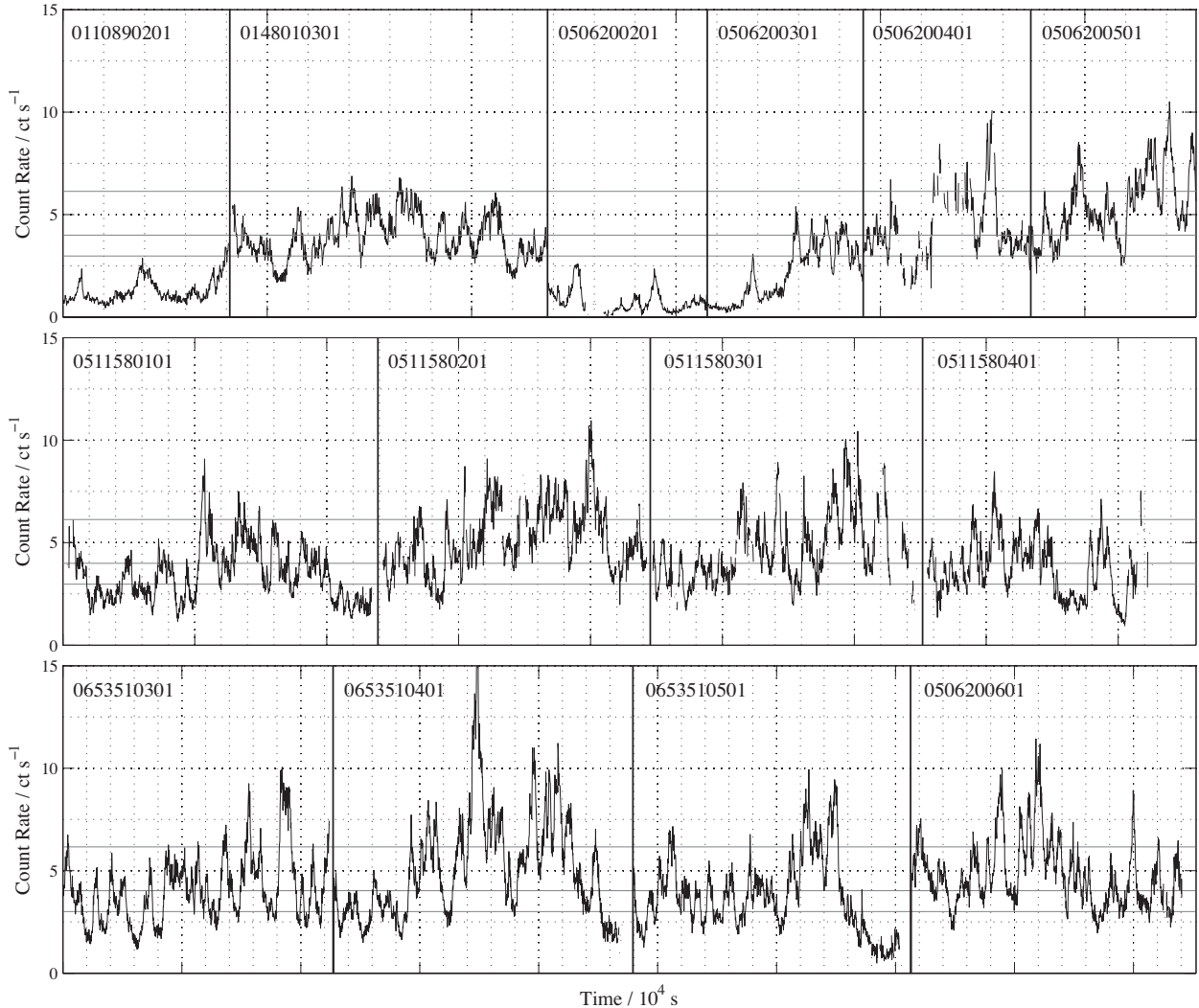}
\caption[]{Background-subtracted, corrected X-ray lightcurves of the of the NLS1 galaxy 1H\,0707$-$495 recorded with the pn detector on board \textit{XMM-Newton} and produced by the SAS task \textsc{epiclccorr}, illustrating the extreme X-ray variability exhibited by this accreting black hole. Vertical lines separate observations (orbits) taking place at different times. Details of the observations can be found in Table~\ref{data.tab}. These light curves were used to construct GTI filters to select cumulative times of varying X-ray flux. Grid lines on the axes are in units of $10^4$\s\ and the solid horizontal lines show the the primary set of independent. non-overlapping flux cuts used in the analysis.}
\label{lightcurves.fig}
\end{figure*}

\section{The Changing X-ray Spectrum}
While there may not be prolonged periods of high and low flux from which lag spectra can be extracted in all sources, it is possible to extract X-ray spectra from the different flux states. Spectra are extracted across varying flux states of the NLS1 galaxy 1H\,0707$-$495. Archival observations of this source between October 2000 and September 2010 total more than 1\Ms\ (Table~\ref{data.tab}), giving long accumulated time periods with a substantial number of counts in each flux state, which is particularly important when using the detailed profile of the relativistically broadened iron K$\alpha$ emission line to constrain the geometry of the corona. The 2011 \textit{XMM-Newton} observation during the period in which 1H\,0707-495 dropped into an extremely low flux state is not included due to the substantially different behaviour of the X-ray spectrum, with little or no directly observed continuum emission detected, though this observation is considered thoroughly by \citet{1h0707_jan11}.

\begin{table}
\centering
\caption{\textit{XMM-Newton} observations from which periods of varying flux were extracted.}
\begin{tabular}{lccl}
  	\hline
   	\textbf{Obs ID} & \textbf{Start} & \textbf{Exposure} & \textbf{Reference} \\
	\hline
	0110890201 & 2000-10-21 & 46.0\ks & \citet{boller+02} \\
	0148010301 & 2002-10-13 & 80.0\ks & \citet{fabian+04} \\
	0506200201 & 2007-05-16 & 40.9\ks & \citet{kara+12} \\
	0506200301 & 2007-05-14 & 41.0\ks & \citet{kara+12} \\
	0506200401 & 2007-07-06 & 42.9\ks & \citet{kara+12} \\
	0506200501 & 2007-06-20 & 46.9\ks & \citet{kara+12} \\
	0511580101 & 2008-01-29 & 123.8\ks & \citet{zoghbi+09} \\
	0511580201 & 2008-01-31 & 123.7\ks & \citet{zoghbi+09} \\
	0511580301 & 2008-02-02 & 122.5\ks &\citet{zoghbi+09} \\
	0511580401 & 2008-02-04 & 121.9\ks & \citet{zoghbi+09} \\
	0653510301 & 2010-09-13 & 116.6\ks & \citet{dauser+12} \\
	0653510401 & 2010-09-15 & 128.2\ks & \citet{dauser+12} \\
	0653510501 & 2010-09-17 & 127.6\ks & \citet{dauser+12} \\
	0653510601 & 2010-09-20 & 129.0\ks & \citet{dauser+12} \\
	\hline
\end{tabular}
\label{data.tab}
\end{table}

\subsection{The Spectrum in Varying Flux States}
X-ray spectra in different flux limits were extracted from observations of 1H\,0707$-$495 using the EPIC pn detector \citep{xmm_strueder} on board \textit{XMM-Newton} \citep{xmm_jansen}. Data collected during individual orbits were reduced separately using the \textit{XMM-Newton Science Analysis System} (SAS) version 12.0.1 using the most recent calibration data for the observations in question.

After initial reduction of the event lists and removal of background flares, the light curves recording the total count rate from the source were extracted (and are shown in Fig.~\ref{lightcurves.fig}) and used to create a \textit{good time interval} (GTI) filter that selects the periods during the orbit in which specified criteria are met. Time intervals were selected in which the total count rate from the source was within specified ranges, shown in Table~\ref{fluxcuts.tab} (the rapidity of the variability and how long is spent at each flux level in the X-ray emission can be seen in Fig.~\ref{lightcurves.fig} showing the light curves from the observations). Relatively broad cuts in flux across all times were selected rather than continuous time periods at different flux levels in order to maximise the number of photon counts within each to enhance spectral analysis.

When studying the variation in parameters between the flux segments and particularly when searching for correlations of these parameters with flux, only non-overlapping segments can be compared such that the different flux cut spectra are composed of independent sets of photons and are therefore statistically independent. The flux segments in Table~\ref{fluxcuts.tab} are grouped into two sets, within which each of the segments are independent and non-overlapping. These groups are treated separately throughout the following analysis.

These GTI filters ere then used to extract the spectra counting only the photons that arrived during the periods in which the count rate was within the required range. The source spectra were extracted from a circular region of the detector centred on the co-ordinates of the point source, 35\asec\ in diameter. Thus, we obtain the average spectrum of the source over all times when the count rate is as required. Corresponding background spectra were extracted from a region of the same size, on the same chip as the source using photons from the same time periods. The spectra were binned using the \textsc{grppha} tool such that there were at least 25 counts in each spectral bin. The photon redistribution matrices (RMF) and ancillary response matrices (ARF) were computed for each spectrum, then the spectra from all orbits in a given flux state were summed under average response matrices.

\begin{figure}
\centering
\includegraphics[width=90mm]{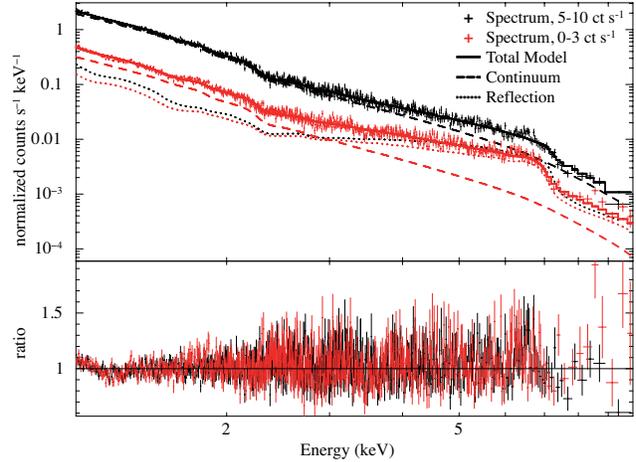}
\caption[]{The EPIC pn X-ray spectrum across the band 1.1-10.0\keV\ of 1H\,0707$-$495 in the lowest (0-3\ctsps) and second highest (5-10\ctsps) flux cuts extracted from the total 1.3\Ms\ dataset. The spectrum between 1.1 and 10.0\keV\ is well described by simply the power law continuum emission from the corona and its relativistically blurred reflection from the accretion disc. The best fitting model components are shown.}
\label{minmaxspec.fig}
\end{figure}

Above 1.1\keV, the X-ray spectrum of 1H\,0707$-$495 is well described as the combination of the continuum emission from the coronal X-ray source (taking the form of a power law) and reflection from the accretion disc. The emissivity profile of this reflection (the reflected flux as a function of position on the disc) takes the form of a twice broken power law \citep{1h0707_emis_paper, miniutti+03} and in the case of an X-ray emitting region extending radially over the disc, the outer break point of this function between the flattened middle region and the outer power law $r^{-3}$ corresponds to the radial extent of the source over the accretion disc \citep{understanding_emis_paper}. Fig.~\ref{minmaxspec.fig} shows the spectrum in high and low flux states with the best fitting model consisting of the continuum emission and its relativistically blurred reflection from the accretion disc. It is immediately apparent that the majority of the variability is due to changes in the continuum with only small changes seen in the reflection spectrum.

In this simultaneous fit to the two flux segments, $\chi^2/N_\mathrm{DoF} = 1.15$, indicating an acceptable, though not perfect fit. The majority of the residuals to the model here adopted lie below 1.5\keV\ and above 7\keV\ (indeed, ignoring the data points below 1.5\keV\ reduces $\chi^2/N_\mathrm{DoF}$ to 1.03). \citet{blustin} find a complex series of emission and absorption features in high resolution spectra below 2.5\keV\ attributed to the narrow cores of atomic features arising from the outer parts of the accretion disc. \citet{dauser+12} conduct a detailed analysis of the most recent \textit{XMM-Newton} observations of 1H\,0707$-$495 and find that the absorption structure seen above 7\keV\ is explained by a complex ionisation structure to the disc, represented in their models by two cospatial reflection components with different ionisation parameters, indicative of a complex ionisation structure to the accretion disc, while small structures between 2 and 5\keV\ can be attributed to a fast outflow from the accretion disc. These features, however, represent only small corrections to the overall shape of the continuum spectrum and the profile of the relativistically broadened iron K$\alpha$ line which are used here to probe the behaviour of the variable X-ray emitting corona, so will not be discussed further in this work in the interests of adopting a simpler model of the X-ray emission and its reflection from which the variations in the corona can be understood more easily.

\begin{table}
\centering
\caption{Flux cut segments for which spectra were extracted from the observations of 1H\,0707$-$495. The mean count rate was calculated from the light curve during each of the cuts and the error corresponds to the standard deviation of the rate within the cut. The segments are divided into two independent groups that should be analysed separately during analysis and when searching for correlations, since each spectrum in each group is constructed from non-overlapping time segments. Also shown is the total effective exposure time selected by each segment and the total number of counts detected therein.}
\begin{tabular}{llll}
  	\hline
   	\textbf{Segment} & \textbf{Mean Count Rate} & \textbf{Exposure} & \textbf{Counts}\\
	\hline
	0 -- 3\ctsps & $1.89 \pm 0.75$\ctsps & $3.94 \times 10^{5}$\s &$1.0 \times 10^6$ \\
	3 -- 4\ctsps & $3.48 \pm 0.26$\ctsps & $1.90 \times 10^{5}$\s & $8.9\times 10^5$ \\
	4 -- 6\ctsps & $4.81 \pm 0.53$\ctsps & $1.90 \times 10^{5}$\s & $1.2\times 10^6$ \\
	6 -- 10\ctsps & $7.04 \pm 0.84$\ctsps & $5.64 \times 10^{4}$\s & $5.4\times 10^5$ \\
	\hline
	0 -- 4\ctsps & $2.43 \pm 0.97$\ctsps & $6.08 \times 10^{5}$\s & $2.0 \times 10^6$ \\
	4 -- 5\ctsps & $4.46 \pm 0.25$\ctsps & $1.18 \times 10^{5}$\s & $7.1\times 10^5$ \\
	5 -- 10\ctsps & $6.20 \pm 0.98$\ctsps & $1.25 \times 10^{5}$\s & $1.4\times 10^6$ \\	
	\hline
\end{tabular}
\label{fluxcuts.tab}
\end{table}

\subsection{The Extent of the X-ray Source}

The best way to determine the change in the extent of the X-ray source as the flux received from the source varies would be to determine the emissivity profile of the reflection component from each flux cut by decomposing the detected iron K$\alpha$ emission line into the contributions from successive radii following the procedure of \citet{1h0707_emis_paper}.  The effective exposures, however, of the spectra in each of the flux cuts are only around 100\ks. This means that, particularly in the lower flux states, there were not a sufficient number of photons detected in the reflection component to properly constrain the emissivity profile in this way, with a large number of degrees of freedom.

Rather than fitting directly for the emissivity profile of the accretion disc, we are guided by the emissivity profiles obtained from the total observations of 1H\,0707$-$495 \citep{1h0707_emis_paper} as well as that of IRAS\,13224$-$3809 \citep{iras_fix} to assume that the emissivity profile takes the form of a twice broken power law and we fit the slopes of the three parts of the profile as well as the locations of the break radii, reducing the number of free parameters in the emissivity profile from 35 (the number of distinct annuli available in the \textsc{kdblur} convolution kernel used in the radial decomposition of the reflection spectrum) to five. The reflection spectrum as measured in the rest frame of the material in the accretion disc was computed by the \textsc{reflionx} model of \citet{ross_fabian} and was convolved with the profile of a single emission line broadened by the relativistic effects from an orbiting accretion disc in the Kerr spacetime using a twice-broken power law form of the emissivity profile, computed using the \textsc{kdblur3} model of \citet{1h0707_emis_paper} constructed as an analytic approximation to the observed emissivity profile that was determined directly and found therein to best reproduce the observe profile of the broad iron K$\alpha$ line. The directly observed continuum emission from the corona was modelled as a power law, giving the total model spectrum
\begin{equation}
	\mathrm{powerlaw} + \mathrm{kdblur3} \otimes \mathrm{reflionx}
\end{equation}
The inclination of the accretion disc and the iron abundance were fixed at the previously found best-fit values of \citet{zoghbi+09} as these certainly do not change during the observations. In the interests of limiting the number of free parameters while obtaining the accretion disc emissivity profile, the ionisation parameter in the \textsc{reflionx} model was also assumed to remain constant between the high and low flux periods within the observations. While in reality, variation in the incident X-ray flux will likely change the ionisation state of the reflecting material, the value obtained in fitting to the whole observation will be the average ionisation parameter during this period. The frozen model parameters are detailed in Table~\ref{par.tab}.

It can readily be shown in this model that the assumed value of the ionisation parameter does not influence the measured location of the outer break radius in the emissivity profiles long as $\xi < 150$\ergcmps. In this `low ionisation' regime, the spectrum is dominated by the lesser-ionised species which produce the prominent iron K$\alpha$ line at 6.4\keV. Once convolved with the relativistic blurring kernel, it is this emission line that provides the strongest measure of the emissivity profile. It is only once the ionisation parameter exceeds 150\ergcmps\ that the measurement of the emissivity profile is affected. Initially, the greater abundance of more ionised species with a vacancy in the L shell causes the iron K$\alpha$ photons to be reabsorbed, weakening the emission line and hence the statistical constraint on the emissivity profile. Finally, once helium-like and hydrogenic iron become prevalent when $\xi > 500$\ergcmps, the K$\alpha$ line is shifted from 6.4\keV\ to 6.67\keV\ and 6.97\keV, respectively and the correct emissivity profile will not be measured when assuming too low an ionisation parameter. As such, accurate determination of the ionisation state of the accretion disc only starts to become important when $\xi > 150$\ergcmps\ and becomes critical when $\xi > 500$\ergcmps.

In the case of 1H\,0707$-$495, the ionisation parameter was not found to increase above 65\ergcmps\ in the highest flux segment (when fitting to the full 0.3-10\keV\ energy range to include the structure in the reflection spectrum below 1\keV), thus the assumption of a constant ionisation parameter does not influence the measured emissivity profile of the accretion disc.

\begin{table}
\caption{Values of frozen parameters in the model that was fit to the spectra of 1H\,0707$-$495 in varying flux states.}
\begin{tabular}{lll}
	\hline
   	\textbf{Component} & \textbf{Parameter} & \textbf{Value} \\
	\hline
	\textsc{kdblur3} & Inclination, $i$ & $53.96\deg$ \\
	\hline
	\textsc{reflionx} & Incident photon index, $\Gamma$ & = \textsc{powerlaw:}$\Gamma$ \\
	& Iron abundance, $A_\mathrm{Fe}$ & 8.88x Solar \\
	& Ionisation parameter, $\xi$ & 53.44\ergcmps \\
	& Redshift, $z$ & $4.10\times 10^{-2}$ \\
	\hline
\end{tabular}
\label{par.tab}
\end{table}

The slope of the power law continuum and the normalisations of the continuum and reflection are fit as free parameters to the spectrum and the slopes and break radii of the emissivity profile are fit, but constrained to be within reasonable ranges to give the expected shape of the emissivity profile (\textit{i.e.} a steep decline, followed by a flatter region before a slope close to $r^{-3}$ over the outer parts of the disc) as shown in Table~\ref{fluxcuts_kdblur3fit.tab}.

\begin{table}
\centering
\caption{Allowed ranges for the parameters of the twice-broken power law emissivity profile when fitting to find the extent of the X-ray source.}
\begin{tabular}{llll}
  	\hline
   	\textbf{Parameter} & \textbf{Fit Range} \\
	\hline
	Index 1 	& 	5 -- 10	\\
	Break radius 1 	& 	3 -- 5\rg 	\\
	Index 2 	&	0 -- 2  \\
	Break radius 2	&	5 -- 35\rg\\
	Index 3		&	2 -- 4 \\
	\hline
\end{tabular}
\label{fluxcuts_kdblur3fit.tab}
\end{table}

\label{expanding_corona.sec}
Fitting the emissivity profile to the observed spectrum (namely the profile of the relativistically broadened iron K$\alpha$ emission line) as a twice-broken power law, the location of the outermost break radius (between the flat part of the emissivity profile and the approximate inverse-cube profile over the outer part of the disc) as a function of count rate for 1H\,0707$-$495 for the first set of four statistically independent, non-overlapping lux segments is shown in Fig.~\ref{fluxcuts_spec.fig:rbreak}. The same increase is seen through the second set, with the break radius increasing from $24.4_{-1.6}^{+3.1}$\rg\ in the lowest of the flux segments to lower and upper limits of $>24.6$\rg\  and $<29.3$\rg\ respectively, in the greater flux segments.

In each of the spectra, the innermost power law index of the emissivity profile was found to be steep (between 7 and 8 in each case), the power law index of the middle section is close to zero in each case, and over the outer part of the disc, the emissivity profile falls off with a power law index around 3.3, consistent with previous findings from spectra averaged over the whole of the observations \citep{1h0707_emis_paper}. These parameters were, however, still allowed to be free in the fitting procedure as any slight variation here that is not accounted for can lead to errors in the determined break radii to compensate for the real shape of the emissivity profile.

In each case, the model was found to provide a good fit to the spectrum, yielding values of the reduced $\chi^2$ fit statistic between 1.0 and 1.1, except in the case of the lowest count rate segments where the value increased to between 1.20 and 1.25. We find, however, that it is the structure in the spectrum below 2.5\keV\ that largely contributes to this as it is not thoroughly accounted for in the simple model we employ here as discussed above. This does not, however, affect our conclusions based upon the profile of the iron K$\alpha$ emission line and slope of the continuum spectrum.

\citet{understanding_emis_paper} demonstrate that in order to reproduce \textit{both} the steep inner part of the emissivity profile \textit{and} the outer break radii that are found, an X-ray emitting corona extending radially at a low height (but with a finite vertical extent that may be allowed to vary, although the emissivity profile is not sensitive to the vertical extent of such a source) above the accretion disc is required, thus alternative models such as a `lamppost' point source or collimated vertical source along the rotation axis of the black hole will not be considered further.

Identifying the outer break radius of the emissivity profile with the outermost radial extent of the X-ray source in a plane parallel to the accretion disc, following \citet{understanding_emis_paper}, we see evidence that the X-ray emitting corona expands as the luminosity increases then contracts as the luminosity decreases again (note we are finding the average extent of the source between all times of a given count rate rather than following the evolution of the source in time as the count rate varies). The Spearman rank correlation co-efficient for the first set of independent flux segments is $\rho=0.84$, indicating $p<0.01$ that the appearance of a correlation is merely due to random chance. A linear relation can be fit to the best-fitting outer break radius for the first set of four flux segments data with gradient $2.1\pm1.3$ ($\chi^2/N_\mathrm{DoF}=0.34$), confirming that the outer break radius is increasing with count rate at the 90 per cent confidence level.

The increase in radius of the corona is slight at 25 to 30 per cent with, at most, a $2\sigma$ variation between points at low and high count rates.

\subsection{The Continuum Spectrum}
\begin{figure*}
\centering
\subfigure[Emissivity profile outer break radius] {
\includegraphics[width=85mm]{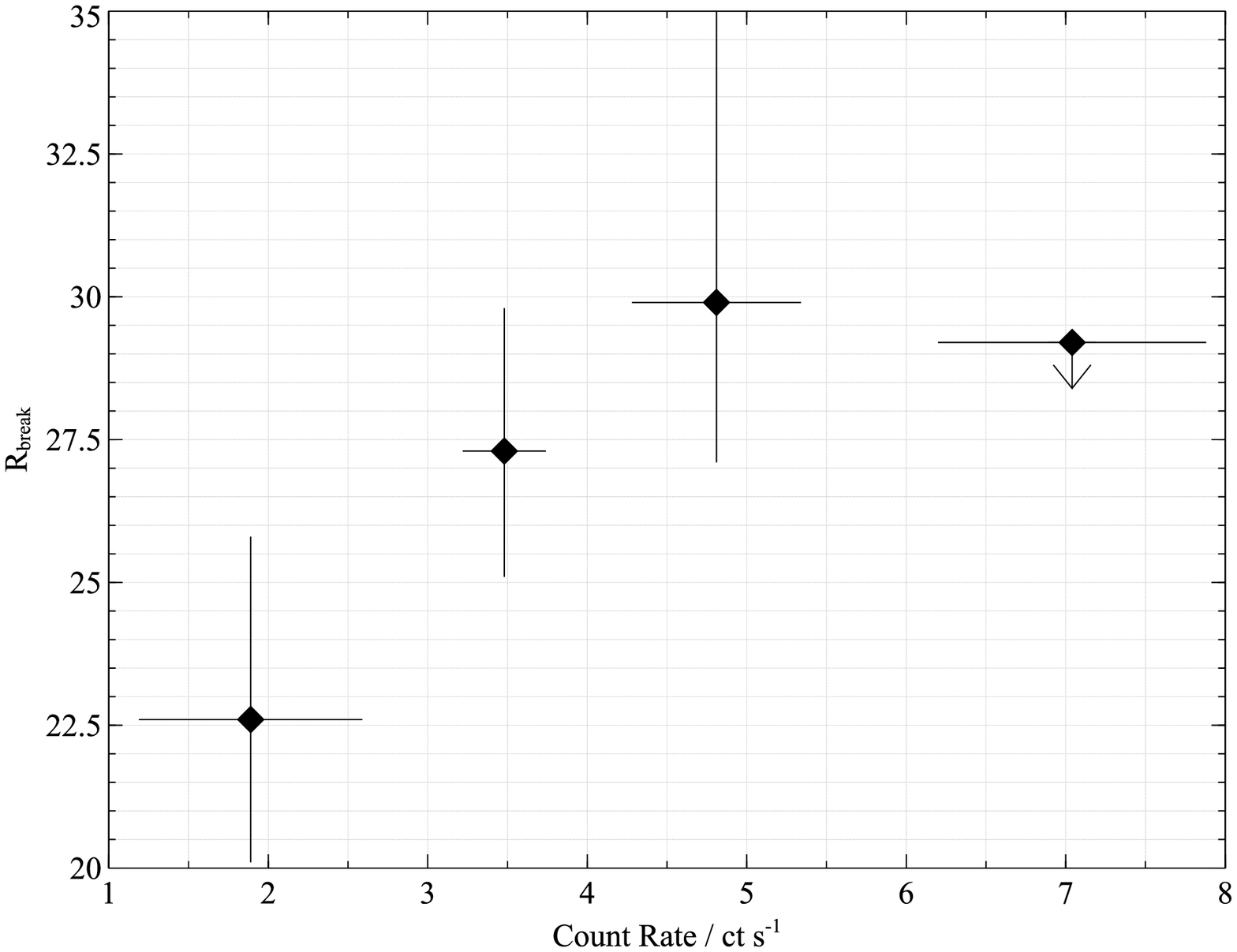}
\label{fluxcuts_spec.fig:rbreak}
}
\subfigure[Photon index of X-ray continuum] {
\includegraphics[width=85mm]{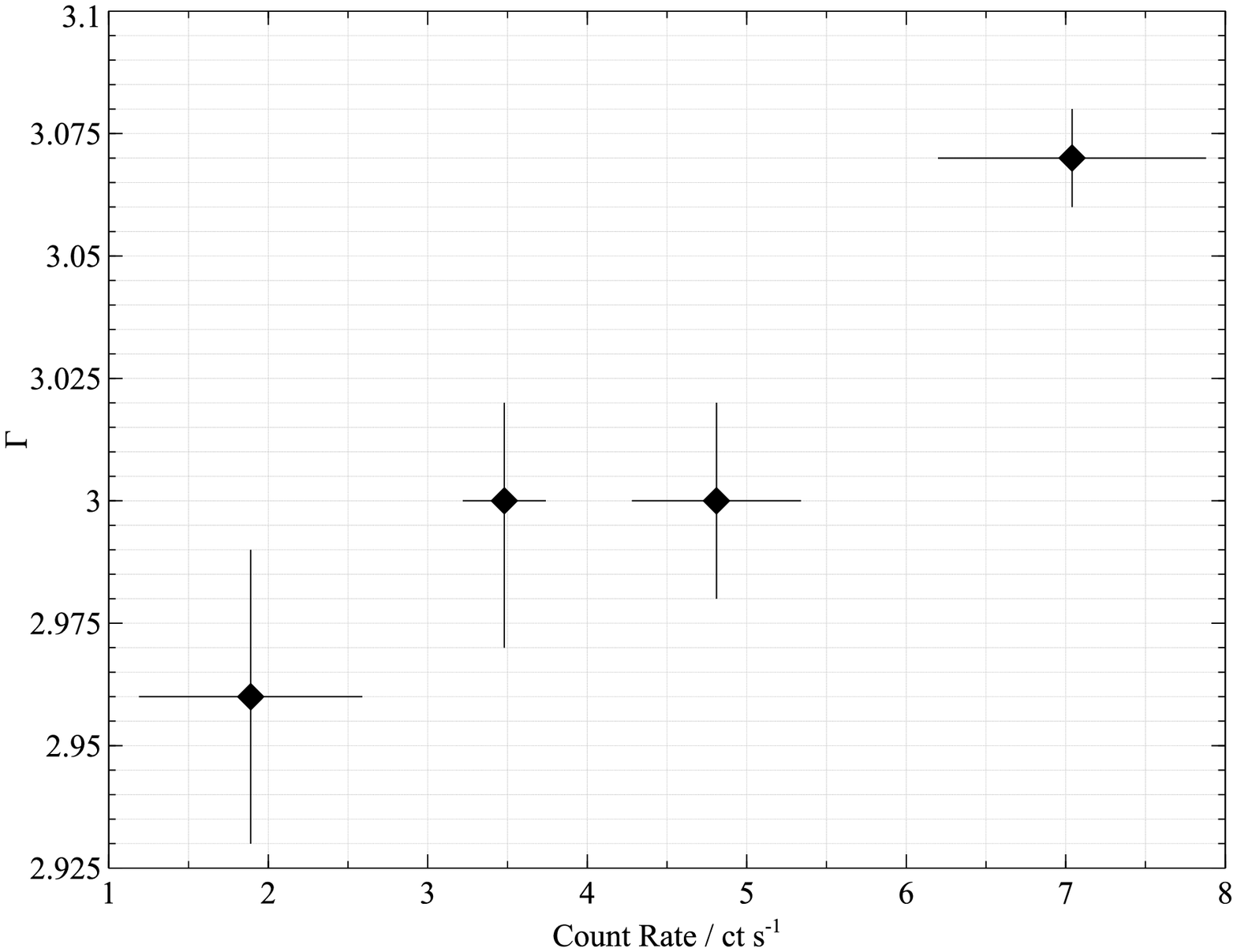}
\label{fluxcuts_spec.fig:gamma}
}
\caption[]{\subref{fluxcuts_spec.fig:rbreak} The outer break radius of the emissivity profile (taken to be a twice-broken power law) of the X-ray reflection from the accretion disc and \subref{fluxcuts_spec.fig:gamma} the photon index of the continuum emission from the corona in 1H\,0707$-$495 as a function of the total count rate from the source for the first set of four non-overlapping, statistically independent flux segments. Error bars correspond to $1\sigma$ on the break radius and spectral index and to the standard deviation of the count rate within the filter segments.}
\label{fluxcuts_spec.fig}
\end{figure*}

Turning to the photon index of the directly observed continuum emission from the corona, $\Gamma$ (where the continuum spectrum takes the form $I_E(E)\propto E^{-\alpha}$ in terms of the spectral energy density or $N_E(E)\propto E^{-\Gamma}$ in terms of photon count rates, with $\Gamma = 1+\alpha$), the best-fit value for the photon index of the continuum spectrum for each flux cut is shown in Fig.~\ref{fluxcuts_spec.fig:gamma}. Experimenting with various energy bands reveals that the best-fit power law index is driven by the continuum-dominated 2-3\keV\ energy band rather than being an artefact of variability in the reflection component or even the thermal emission from the disc just encroaching on the region of the spectrum immediately above 1.1\keV\ and consistent results are obtained if the lower energy bound is increased to either 1.5 or 2.0\keV. 

It is clear to see that as the luminosity of the X-ray source increases, the coronal emission becomes softer, with a more steeply falling spectrum such that fewer hard X-rays are emitted compared to the softer photons. In this case, the Spearman rank correlation co-efficient for the first set of flux segments is $\rho=0.95$, indicating $p<0.001$ that the appearance of a correlation is merely due to random chance. In this case, the best-fit linear relation is better constrained with gradient $0.02\pm0.002$ ($\chi^2/N_\mathrm{DoF}=0.74$).

The same increase is seen in the second set of flux segments, with the photon index increasing from $2.97_{-0.02}^{+0.01}$ to $2.99_{-0.02}^{+0.03}$ and $3.03_{-0.01}^{+0.01}$ as the flux increases.

This trend is well-known in the X-ray emission in AGN. As the luminosity increases, the X-ray continuum spectrum becomes softer \citep[\textit{e.g.},][]{mark_edel_vaughan}.

\section{The Reflected vs. Direct Continuum Flux}
\subsection{Observations}
A key prediction of models in which the increase in luminosity of an X-ray emitting corona corresponds to the spatial expansion of that corona is that a collapse of the corona into a more confined region around the central black hole not only leads to reduced overall luminosity but also must result in an increase in the number of photons hitting the disc to be reflected relative to the number that are able to escape to be observed in the continuum \citep[\textit{e.g.}][]{miniutti+04,fukumura+07}. This increase in the reflection fraction is due to emission from closer to the black hole being bent towards the black hole and thus focussed onto the accretion disc and was invoked to explain the simultaneous steepening of the iron line emissivity profile and disappearance of the continuum when 1H\,0707$-$495 dropped into an extremely low flux state in January 2011 \citep{1h0707_jan11}.

The fraction of the photons that are reflected and observed directly from a spatially expanding corona can be explored through general relativistic ray tracing simulations. Ray tracing simulations will predict the number of photons emitted from the corona that hit the disc compared to the number that escape to infinity. They will not, however, directly predict the number of photons that will be \textit{observed} in the reflection component of the observed spectrum without inputting a detailed model of the (spatially resolved) properties of the disc material and the `reflection' processes that take place when photons are incident upon it (Compton scattering, photoelectric absorption, fluorescent line emission, bremsstrahlung, \textit{etc.}). As such, the number of photons that are seen in the reflection spectrum is not directly indicative of the number of photons that hit the disc and cannot be directly compared to the number of photons detected in the continuum.

It is, however, possible to probe the \textit{variation} in reflection fraction by measuring the photon flux seen in the reflection component of the spectrum \textit{as a function of} the photon flux directly detected in continuum emission from the corona. This is achieved by fitting the model consisting of a power law continuum and reflection from the accretion disc described by the \textsc{reflionx} code as detailed above and then computing the flux (which can be found as both the photon flux and energy flux) that would be detected by the telescope in each of these spectral model components by folding the fitted model through the instrument responses. When considering reflection from the accretion disc, significant flux emerges in the `Compton hump' around 30\keV\ and in emission lines at soft X-ray energies, around 0.1\keV. It is therefore necessary to compute fluxes integrating photons over a wide energy range, here taken to be 0.1-100\keV\ and the instrument response is extrapolated from its limit at 12\keV\ up to 100\keV\ for the purposes of computing the flux represented by the model components.

The continuum and reflected fluxes are determined by fitting a model spectrum to the data over the 0.3-10\keV\ energy band (the full energy range of the pn detector). Photoelectric absorption by Galactic material is accounted for in the model fit over the 0.3-10\keV\ energy band, using the \textsc{phabs} model in \textsc{xspec}, so too is the thermal emission detected from the accretion disc, modelled by a black body spectrum whose temperature is around 0.05\keV, consistent with the model of \citet{zoghbi+09} for this source. The ionisation parameter is allowed to vary as a free parameter in this instance as changes in the ionisation state of the disc could affect the overall flux as changes in ionisation alter the large number of emission lines below 1\keV\ in which a significant part of the reflected flux emerges \citep[\textit{e.g.}][]{ross_fabian}, though, as already discussed, does not affect the measurement of the accretion disc emissivity profile from the 6.4\keV\ iron K$\alpha$ line. It is therefore necessary to include the 0.3-1.1\keV\ energy range to ensure that the ionisation parameter is properly determined and find that excluding the 0.3-1.1\keV\ part of the spectrum causes the ionisation parameter to be systematically underestimated, being measured as low as half the value obtained using the full 0.3-10\keV\ energy band. When fitting our model assuming a single ionisation parameter over the whole disc, the ionisation parameter is found to increase systematically from 59\ergcmps\ in the lowest flux segment to 65\ergcmps\ in the highest flux segment, thus the earlier assumption that the disc remains in the `low ionisation' state and, as such, the measured emissivity profile is not altered by the ionisation state of the disc is justified.

\begin{figure}
\centering
\includegraphics[width=85mm]{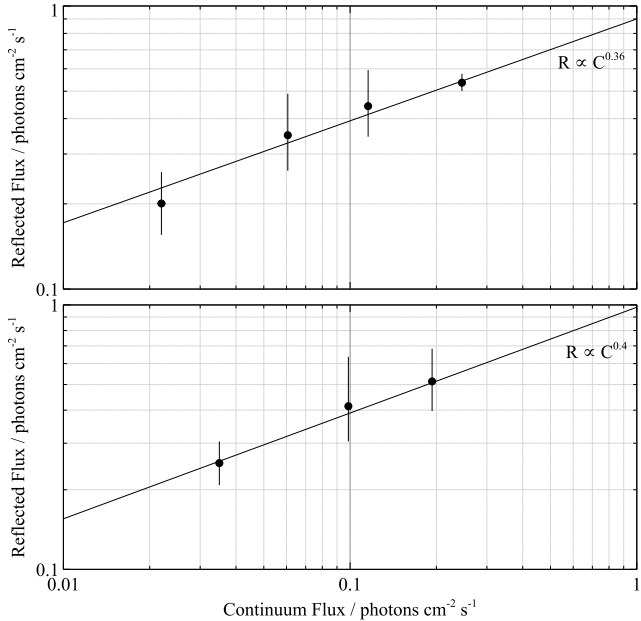}
\caption[]{The relationship between the photon counts in the X-ray reflection from the accretion disc and the directly observed continuum emission in 1H\,0707$-$495 obtained by fitting a model consisting of a power law continuum and reflection from the accretion disc described by the \textsc{reflionx} model of \citet{ross_fabian} as well as Galactic absorption and black body emission from the accretion disc to the energy band 0.3-10\keV, along with the best fitting power laws. The upper panel shows the first set of independent, non-overlapping, flux segments and the lower panel, the second set. The best fit relation is found separately for each set.}
\label{refplflux.fig}
\end{figure}

Results are shown in Fig.~\ref{refplflux.fig} and we approximate the relationship between the, the reflected flux, $R$, and the continuum flux, $C$, as a power law with $R\propto C^\beta$. Considering just the first set of four independent flux segments, we find for 1H\,0707$-$495 that the index of this power law, $\beta = 0.35_{-0.06}^{+0.14}$ which is consistent with that found when considering the second set of independent flux segments, for which $\beta = 0.40_{-0.15}^{+0.20}$. Critically, we find that $\beta < 1$, as we shall discuss in the forthcoming section.

\subsection{Ray Tracing Simulations of the Reflection Fraction from Extended Coron\ae}

If the reflection fraction remained constant (\textit{i.e.} the source geometry remained constant or reflection took place from distant material and experienced no gravitational light bending that enhances reflection as the X-ray source changes) and the reflected flux changed only due to the variations in the illuminating luminosity (the intrinsic variability in the source), one would expect the reflected photon flux to rise linearly with the continuum flux detected directly from the corona. This linear relation would hold no matter how the intrinsic luminosity of the source varies in time. At the other extreme, if the total photon count rate emitted from the source remains constant but its location or geometry changes, changing the fraction of the emitted photons that are reflected, the reflected photon flux would fall linearly as the continuum flux increases (\textit{i.e.} the shape $y=1-x$ where $y$ and $x$ correspond respectively to the reflected and continuum flux). The total number of photons is conserved, they are just moved from one component to the other, except when the source is close to the black hole, when both the reflected and continuum fluxes would fall as a significant proportion of the photons emitted from the corona are lost through the event horizon.

If the luminosity of the source is allowed to vary while it expands, the relationship between the continuum and reflected fluxes cannot be simply interpreted in terms of the extent or position of the corona. Instead, a simplified model is constructed to test if the data we obtain are consistent with the picture of an expanding corona giving rise to increased X-ray luminosity. While we do not know how the intrinsic luminosity of the X-ray source changes, we are guided by the observation that the radial extent of the source, $r$ increases as the source becomes brighter, to approximate the total luminosity as a power law in the radius of the emitting region, $L\propto r^{\beta}$. The vertical extent of the source, $\Delta z$ must also increase if the reverberation lag time between the continuum emission and its reflection from the accretion disc is to lengthen in the high flux state, though this variation is not a necessity to describe these data.

Ray tracing simulations are used to compute the fraction of the emitted photons that hit the accretion disc, are able to escape to be observed as the continuum and that are lost into the black hole event horizon as a function of the source radius. The vertical extent of the source is either held constant or defined to increase linearly with the radius within the limits of the coronal geometry inferred from emissivity and reverberation lag measurements (though it turns out that the functional form of reflection fraction is not greatly altered by how the vertical extent of the source varies with radius). The general relativistic ray tracing algorithm of \citet{understanding_emis_paper} is employed. The X-ray emission from the corona is sampled using a Monte Carlo technique, starting rays in random directions at random locations within the defined region of the corona. They are traced by numerical integration of the null geodesic equations until they either reach the equatorial plane (defined to be the accretion disc), are lost within the black hole event horizon or the innermost stable circular orbit (the inner edge of the accretion disc) or escape to a limiting radius of 10,000\rg, at which point they are said to be able to be detected as part of the X-ray continuum.

These simulations are similar in their aim to those of \citet{miniutti+04} and \citet{fukumura+07} who calculate the reflection fractions from an accretion disc illuminated by an isotropic point source of radiation located on the rotation axis of the black hole. \citet{fukumura+07} compute the radiative transfer analytically, however here we numerically integrate the geodesic equations describing each ray and extend the calculation to a generalised, spatially extended X-ray source by starting rays travelling in random directions at random locations within the defined bounds of the corona (thus simulating an optically thin corona). Full details of the ray tracing algorithm can be found in \citet{understanding_emis_paper}.

It is assumed that the reflected flux varies proportional to the number of photons incident upon the accretion disc (\textit{i.e.} regardless of the scattering, absorption and re-emission processes that give rise to the X-ray reflection spectrum, doubling the flux incident upon the disc doubles the reflected flux). The fractions of photons incident upon the disc and able to escape to become part of the continuum are multiplied  by the total luminosity of the source using the above parametrisation, $L\propto r^{\beta}$. Note that here, $L$ represents the integrated luminosity of the corona and its dependence on the radial extent of the corona, $r$, rather than a luminosity profile as a function of radius within the corona.

\begin{figure*}
\centering
\begin{minipage}{175mm}
\subfigure[Constant photon count]{
\includegraphics[width=56mm]{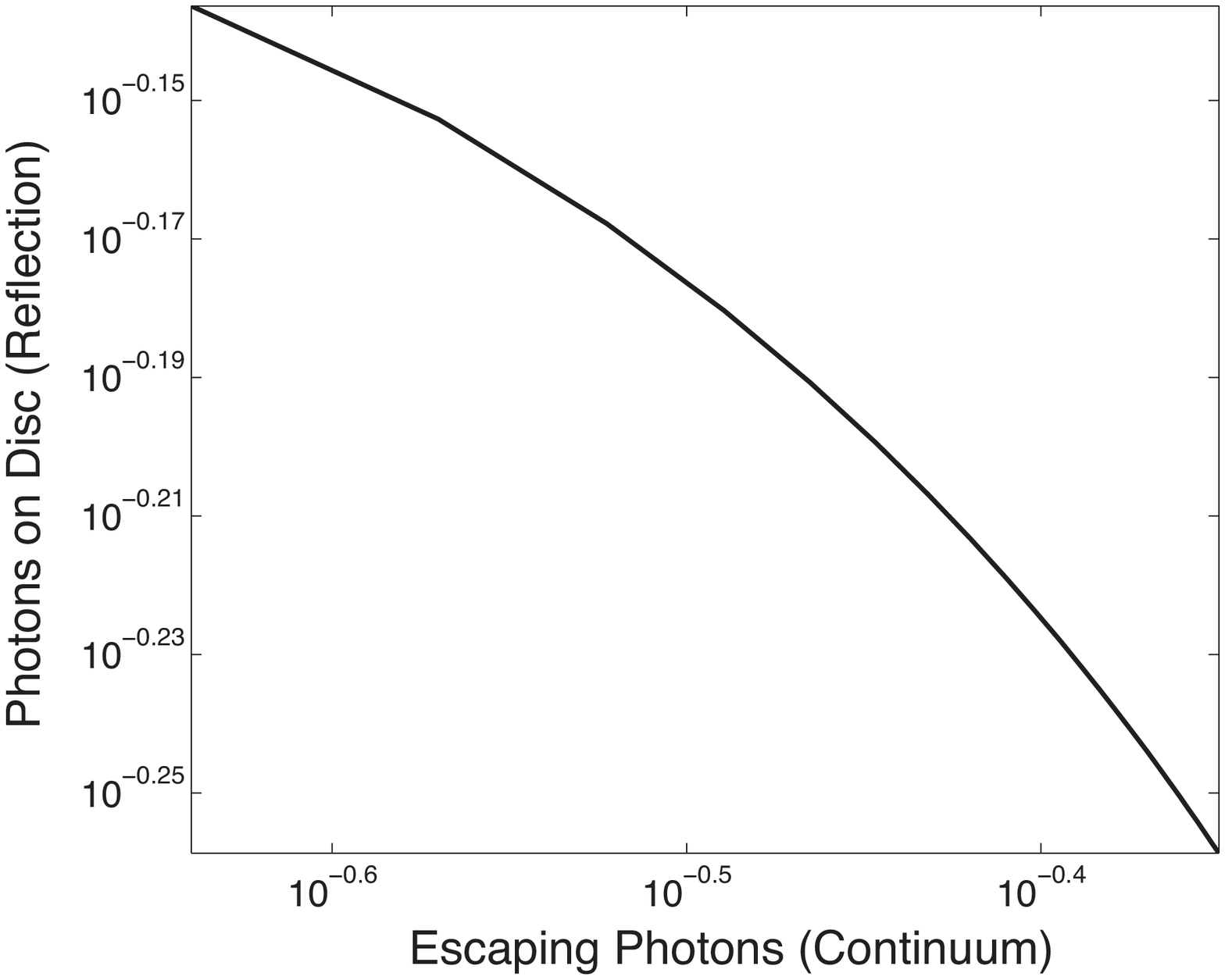}
\label{refpl_theory.fig:const}
}
\subfigure[$L\propto r^2\,\Delta z$]{
\includegraphics[width=56mm]{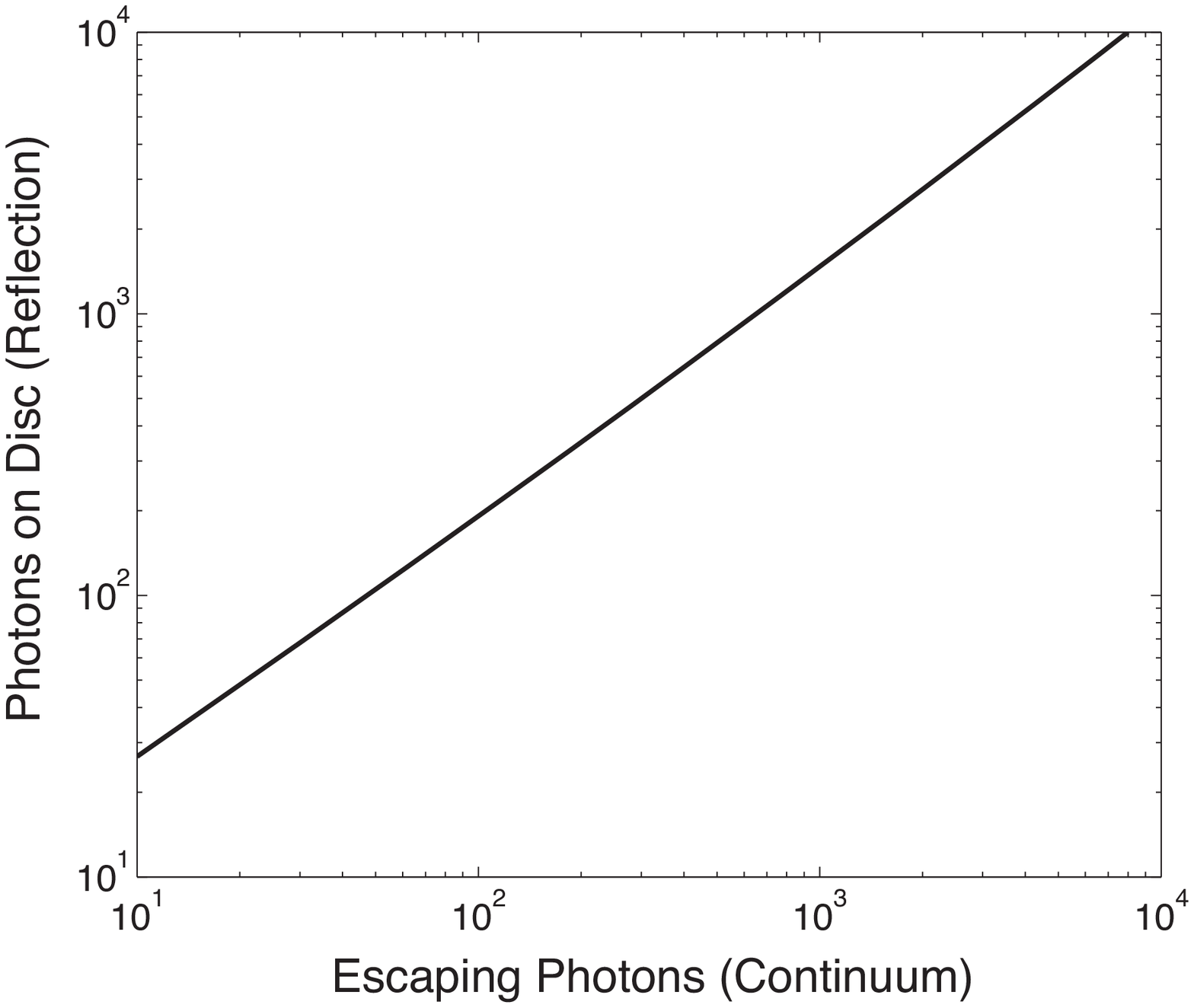}
\label{refpl_theory.fig:vol}
}
\subfigure[$L\propto r^{0.35}$]{
\includegraphics[width=56mm]{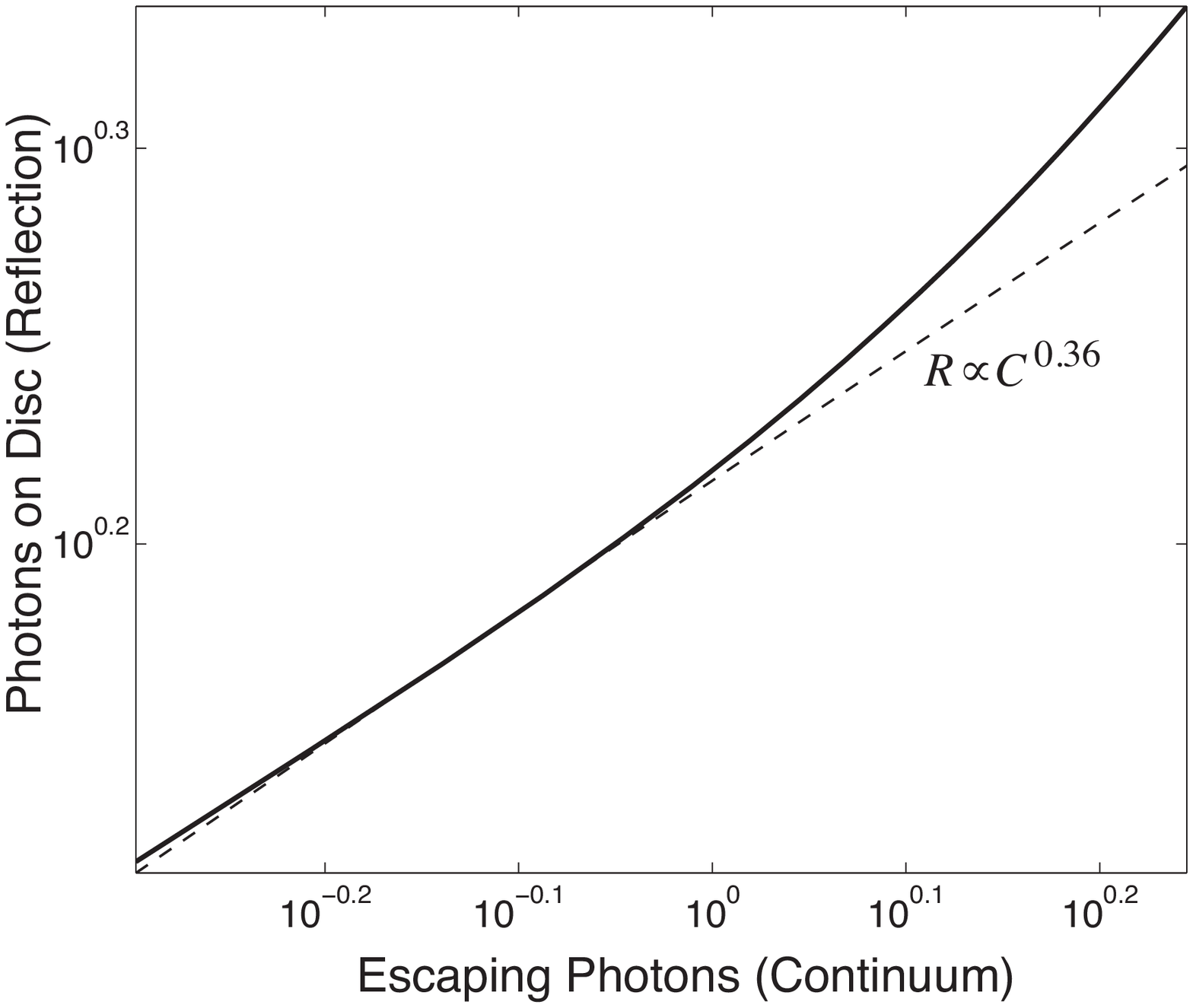}
\label{refpl_theory.fig:r035}
}
\caption[]{The relationship between the photon counts in the X-ray reflection from the accretion disc and the directly observed continuum emission computed from ray tracing simulations counting the rays that hit the disc and that are able to escape to infinity. The coron\ae\ extend radially from 2 to 50\rg\ and vertical extents vary linearly with the radial extent from 0.5\rg\ to 5\rg\, with the base of the corona fixed at 1\rg, for \subref{refpl_theory.fig:const} constant source luminosity where the total photon count is constant and photons are just shifted between the two groups giving a $y=1-x$ relationship, except for the most confined sources where a substantial fraction of the photons is lost into the black hole. In \subref{refpl_theory.fig:vol}, the source luminosity varies proportional to its volume, $r^2\,\Delta z$. In this case the significant intrinsic variability outweighs the change in reflection fraction resulting in the reflected flux rising linearly with the continuum flux. Finally, in \subref{refpl_theory.fig:r035}, the intrinsic, total luminosity of the source is taken to vary as $r^{0.35}$ to reproduce the $R\propto C^{0.36}$ relationship seen in 1H\,0707$-$495 over the source radii up to around 30\rg. The total luminosity of the corona is increasing as it expands through at an arbitrary rate, not directly in proportion to its volume. At larger source radii, the reflection fraction becomes approximately constant and the reflected flux starts to rise linearly with the continuum flux. The absolute scaling on each axis is arbitrary; the power law relationship between the two is the relevant quantity.}
\label{refpl_theory.fig}
\end{minipage}
\end{figure*}

Fig.~\ref{refpl_theory.fig} shows the predicted relationship between the observed continuum, $C$, and reflected flux, $R$, for three simple cases. First is the case of a corona whose total luminosity remains constant but which expands spatially, shown in Fig.~\ref{refpl_theory.fig:const}. When the corona is smaller (and less luminous), more of the photons emitted are focussed towards the black hole and hence on to the accretion disc as more of the corona is confined in the stronger gravitational field close to the black hole. This increase in the fraction of the radiation that is reflected compared to that able to escape to be observed directly in the X-ray continuum leads to the $y=1-x$ relationship as discussed above. This is except for the smallest coron\ae\ where the reflected fraction drops owing to a significant number of the emitted photons being lost through the event horizon or within the innermost stable orbit, missing the disc. This particular case is an extension of the model of \citet{miniutti+04} who compute the reflected \textit{vs.} continuum flux for a constant-luminosity point source that is moved vertically above the black hole and accretion disc and shows the same trend at large heights but a more extreme drop in reflected flux at the low continuum fluxes when the source is at a low height and photons are lost into the black hole.

When the total luminosity of the corona simply increase proportional to its volume, Fig.~\ref{refpl_theory.fig:vol}, the reflected flux is seen to rise linearly with the directly observed continuum flux. In this case, the increase in total luminosity overcomes the decrease in the fraction that is reflected as the corona expands.

We find that in order to produce the observed flux-flux relation for 1H\,0707$-$495, $R\propto C^{0.36}$, the total count rate emitted from the entire volume of the corona obeys $L\propto r^{0.35}$ (Fig.~\ref{refpl_theory.fig}) in this simple model, shown in Fig.~\ref{refpl_theory.fig:r035}. We find that this is insensitive to how the vertical extent of the source varies with radius, so long as it does not rise more rapidly than linearly with the source radius.

\section{Discussion}
\subsection{The Changing Extent of the Corona}
Identifying the outer break radius of the emissivity profile with the outermost radial extent of the X-ray source in a plane parallel to the accretion disc, following \citet{understanding_emis_paper}, we see evidence that the X-ray emitting corona expands as the luminosity increases then contracts as the luminosity decreases again. \citet{understanding_emis_paper} show that the emissivity profile of the accretion disc is sensitive to the radial extent of the X-ray emitting region over the plane of the accretion disc, while being relatively insensitive to its vertical extent perpendicular to the disc plane in the case of a radially extended source at low height, as is required to match the observed form of the emissivity profile. The emissivity profile only becomes sensitive to the vertical extent of the corona once this becomes greater than the radial extent, though such a geometry is not able to simultaneously reproduce the location of the outer break radius and the steep inner part of the emissivity profile \citep{dauser+13,understanding_emis_paper}.

\citet{lag_spectra_paper} show that varying the radial extent of the X-ray emitting region above the accretion disc has only a slight effect on the measured reverberation lag seen in the reflection from the accretion disc. In fact, increasing the radial extent of the source decreased the reverberation lag since the X-rays emitted further from the central black hole do not experience such an extreme Shapiro delay which slows down their passage from the inner parts of the corona. Therefore, if the increasing lag time with increasing count rate observed by \citet{kara_iras_lags} in IRAS\,13224$-$3809 is to be extrapolated to 1H\,0707$-$495 as a general behaviour, the source must also be expanding vertically (\textit{i.e.} perpendicular to the plane of the accretion disc), though the vertical extent of the corona is not well constrained by the emissivity profile. That said, it is not physically unreasonable to picture a corona of accelerated particles expanding in all directions (either through more particles being accelerated or the accelerated particles being less confined) as more energy is injected into it to increase the luminosity.

\subsection{Variability in the Continuum Spectrum}
Comptonisation of thermal seed photons emitted from the accretion disc by high energy electrons in the corona has been widely explored as the means of producing the X-ray continuum seen from accreting black holes \citep[see, \textit{e.g.},][]{galeev+79,titarchuk-94}. Following \citet{sunyaev_trumper} and modelling the corona as a spherical plasma at temperature $T$, with optical depth $\tau$, at photon energies $\ll E/k_\mathrm{B} T$, the continuum spectrum produced from this corona by the Comptonisation of thermal seed photons is a power law, with specific flux $J_{\nu}\propto \nu^{-\alpha}$, where the spectral index is given by

\begin{align}
\label{compton.equ}
\alpha &= -\frac{3}{2} + \left( \frac{9}{4} + \gamma \right)^\frac{1}{2}
\end{align}
Where, for a spherical geometry,
\begin{align*}
\gamma &= -\frac{\pi^2}{3} + \frac{m_\mathrm{e} c^2}{k_\mathrm{B} T\left(\tau + \frac{2}{3}\right)^2}
\end{align*}

And the photon index, $\Gamma = 1+\alpha$. At photon energies $\gg E/k_\mathrm{B} T$, Comptonisation produces an exponentially decaying spectrum (a `Wein tail'), with the cut-off in the power law occurring at $E_\mathrm{cut} \sim k_\mathrm{B} T$.

A softer continuum spectrum (with a greater photon index) is produced if the average energy of each individual electron in the corona is reduced (that is to say the temperature, $T$ is lower), or if the number density of scattering electrons and hence the optical depth through which the seed photons scatter before they can escape the corona is reduced. Given that we do detect relativistically blurred line emission from the inner parts of the accretion disc that would be beneath the corona we infer from the observed emissivity profile, the optical depth of the corona due to Compton scattering cannot be much above unity as an optically thick corona should scatter the emission we detect from the inner parts of the disc.

In order for the spectrum to soften, the average energy per electron decreases, or the optical depth through the corona is reduced. However, more electrons are accelerated into a corona filling a larger volume. This would provide a larger cross section for scattering seed photons emitted from the accretion disc, explaining the increase in count rate of X-ray photons emitted from the corona.

If the optical depth experienced by the seed photons through the corona is assumed to remain constant, we can calculate the variation in coronal temperature required to produce the observed variation in the photon index of the X-ray continuum using Equation~\ref{compton.equ}. The photon index was found to vary between 2.95 and 3.075. Computing the exact temperature of the corona would require either the energy of the cut-off to the power law continuum spectrum to be measured or the optical depth to be known. However, selecting any value for the optical depth between 0.5 and 5 yields from Equation~\ref{compton.equ} the consistent result that the temperature of the corona is varying by approximately 8 per cent. For instance, taking $\tau=1$ as an upper limit to the optically thin regime in order to give a lower limit to the temperature, we find that $T$ varies between 58 and 63\keV\ to produce the observed range in spectral indices, or for $\tau=0.5$, $T$ ranges from 118\keV\ to 128\keV. It should be noted that Equation~\ref{compton.equ} is derived from the Kompaneets equation \citep[see, \textit{e.g.}][]{lightman+87} which is derived via a diffusion approximation, so is formally valid in the limit $\tau > 1$. Here, we consider the limiting case as a vastly simplified model, to estimate the magnitude of the changes occurring within the corona, though a more thorough treatment is beyond the scope of this work, though the results of \citet{titarchuk-94} for Comptonisation in an optically thin plasma give broadly consistent results for the cases considered here.

Such high coronal temperatures and, indeed, low optical depths are plausible in AGN. For instance, \citet{burlon+2011} find that stacking the spectra of all 199 non-blazar AGN from the \textit{Swift BAT AGN Survey}, covering the energy range 15-195\keV, constrains the cut-off energy to be greater than 80\keV\ and in many cases the data are best modelled fixing the cut-off energy at 300\keV. While measuring the exact energy of the cut-off is out of the reach of present instrumentation, the measurement of both the cut-off energy and any variation therein due to changes in the coronal temperature may be achievable in bright AGN with the \textit{Soft Gamma-Ray Detector} on board the forthcoming \textit{Astro-H} X-ray observatory offering spectral coverage up to 600\keV\ \citep{astroh_sgd}.

\subsection{The Reflected \textit{vs.} Continuum Flux}

Na\"ively, if the average energy density of the corona remained constant and it simply grew, the luminosity of an axisymmetric corona will follow $L\propto r^2 \Delta z$ or $L \propto r^3$ if the height were to increase linearly with the radial extent. This relation remains valid for a corona that produces X-ray emission through the inverse-Compton scattering of seed photons from the accretion disc below, so long as the corona remains optically thin to both the X-rays it produces (which it must be for us to see the characteristic reflection signatures from the inner parts of the accretion disc) and to the seed photons. The scattering cross section for the corona in this case simply increases proportional to its volume, increasing the number of scattered photons accordingly.

Finding such a low power law index relating the luminosity to the radial extent of the source implies that the number of seed photons scattered to form the continuum per unit volume \textit{decreases} as it expands. In the context of Comptonisation by an optically thin corona, this means the average number density of scattering particles (and hence the cross-section per unit volume for interaction with seed photons from the accretion disc below) is reduced. Coupled with the observed softening of the continuum spectrum as the corona is inferred to expand, this portrays a corona whose average temperature and density throughout its extent is decreasing.

\subsection{Structure Within the Corona}

This reduction in photons scattered per unit volume and spectral softening can be explained not only by a wholesale reduction of the energy per particle or number density throughout the entire volume, but also if the corona was the most dense and energetic in its central regions and the expanding and contracting outer parts of the corona were either made up of less energetic particles or were less dense. This is in line with the two component corona model of \citet{taylor_uttley_mchardy} who were able to explain the broadband spectral variability of MCG--6-30-15 by a harder power law continuum component that remained constant with a variable soft power law component driving much of the variability.

If the outer part of the corona were less dense, the optical depth to photons just scattering through these outer parts is less meaning that (disregarding any variation in the temperature of these regions), the spectrum produced would be softer. This can, again, be illustrated using Equation~\ref{compton.equ}. If the coronal temperature is fixed to be $T=100$\keV, the average optical depth to electron scattering of the corona to the seed photons, $\tau$ reduces by around 7 per cent from 0.65 to 0.59 to explain the observed softening in the spectrum from $\Gamma=2.95$ to $\Gamma=3.08$. The result is similar when the temperature is fixed to be 50\keV.

The transient nature of the outer parts of the corona as it expands and contracts mean it is the softer part of the coronal emission that is more variable.

This is most likely an overestimate to the variation in the optical depth, as, of course, changes to both the coronal temperature and optical depth are likely to contribute to the change in spectral slope. We do also approximate the corona to being a spherical cloud with the seed photon source embedded at the centre. Measurements of the coronal extent via the accretion disc emissivity profile suggest an extended structure covering part of the accretion disc (though more of the seed photons will be emitted from the more central parts of the disc, \textit{e.g.} \citealt{novthorne}). Many models too suggest that the X-ray continuum emission arises from ephemeral `flares' of accelerated particles rather than a static corona \citep[\textit{e.g.}][]{beloborodov,merloni_fabian}. In these cases, the values we derive here for the temperature and optical depth represent averages across the corona that would apply to a spherical corona producing the same spectrum and the `corona' represents the volume surrounding the black hole and accretion disc in which these flares take place. The model of \citet{beloborodov} suggests that the reflection of the radiation produced by the flares from the accretion disc causes the particles in the flare to be rapidly accelerated to relativistic velocities away from the disc, through the radiation pressure exerted upon them. The continuum radiation is therefore beamed away from the disc, resulting in a reduced reflection fraction and, in this model, the slope of the continuum spectrum is determined almost entirely by the Lorentz factor of the scattering particles producing the continuum. This model is, however, inconsistent with the typically large reflection fractions, $R/C>1$, seen and, as it is presented, is unable to reproduce the steep continuum spectra, $\Gamma > 2$, characteristic of NLS1 galaxies.

\section{Conclusions}
By analysing spectra obtained for the NLS1 galaxy 1H\,0707$-$495 in states of high and low flux, evidence has been found for the expansion of the X-ray emitting corona as the coronal luminosity increases and contraction to a more confined region around the black hole as the luminosity decreases again. The emissivity profile of the accretion disc, obtained by fitting a twice-broken power law emissivity profile to the relativistically broadened iron K$\alpha$ emission line implies that the radial extent of the corona in 1H\,0707$-$495  over the plane of the accretion disc varies by 25 to 30 per cent as the total flux received varies by a factor of two to three. Combined with measurements of an increasing reverberation lag during prolonged high-flux periods of IRAS\,13224$-$3809 and the low state of 1H\,0707$-$495 discussed in \citet{1h0707_jan11}, a picture is emerging of X-ray emitting coron\ae\ that expand and contract both radially and vertically above the plane of the accretion disc as more or less energy is injected from the accretion flow.

It is observed that as the flux received from the source increases, the continuum spectrum becomes softer, while in order to reproduce the observed relationship between the primary continuum and reflected fluxes by simply following the changing reflection fraction as the corona expands and contracts it is required that the total luminosity of the corona expands only as a weak function of its radius rather than scaling with its volume. These observations suggest that as the corona expands, the energy per scattering particle (the coronal temperature) decreases and so too does the number density and hence the average optical depth experienced by the seed photons through the corona. There being more scattering particles spread through a larger corona increases the total scattering cross section and explains the increase in luminosity if the corona is said to arise from the inverse-Compton scattering of thermal seed photons emitted from the accretion disc.

This analysis demonstrates the understanding that can be gained of the processes at work in the X-ray emitting corona around a black hole through detailed analysis of the observed data combined with insight gained from theoretical predictions. These conclusions are, however, limited to 1H\,0707$-$495 due to the much longer total exposure time available on this source. It will be possible to draw firmer conclusions from longer observations of other objects such that high quality spectra can be obtained in many different states of the systems under consideration. 

Measurements of changes to the X-ray emitting corona, causing the variability we see in the luminosity, shed light on the underlying mechanisms driving the variability and ultimately the means by which energy is liberated from the accretion flow into the corona. Deriving the fundamental behaviours of the corona seen in observational data place important constraints on the theoretical models and simulations of material accreting onto supermassive black holes.

\section*{Acknowledgements}
DRW is supported by a CITA National Fellowship. ACF thanks the Royal Society for support. This work is based upon observations obtained with \textit{XMM-Newton}, an ESA science mission 
with instruments and contributions directly funded by ESA Member States and NASA. We thank the anonymous referee for their useful feedback on the original manuscript.

\bibliographystyle{mnras}
\bibliography{agn}

\label{lastpage}

\end{document}